\documentclass[letterpaper,natbib,twocolumn,10pt]{article}
\usepackage[]{hyperref}
\usepackage{usenix2019_v3}

\usepackage{times} 
\usepackage{amsmath} 
\usepackage{amsfonts} 
\usepackage{amssymb} 
\usepackage{authblk}
\usepackage{bold-extra} 
\usepackage[T1]{fontenc} 
\usepackage{anyfontsize} 
\usepackage[symbol]{footmisc}
\usepackage[scaled]{beramono}
\newcommand{\code}[1]{\texttt{\footnotesize #1}}

\usepackage{listings}
\usepackage{algorithm2e}
\usepackage{algorithmic}

\usepackage{alltt}
\usepackage[small,compact]{titlesec}
\titlespacing*{\section}{0pt}{2pt}{1pt}
\titlespacing*{\subsection}{0pt}{2pt}{1pt}
\titlespacing*{\subsubsection}{0pt}{2pt}{1pt}
\titlespacing*{\paragraph}{0pt}{1.75pt}{3pt}
\usepackage[small,bf]{caption}
\usepackage{subcaption}
\usepackage{tabularx}
\usepackage{wrapfig}
\usepackage{graphicx}
\usepackage{multirow}
\usepackage{blkarray}
\usepackage{balance}
\usepackage{enumitem}
\setlist{nolistsep}

\usepackage{fancyhdr}
\makeatletter
\newcommand\footnoteref[1]{\protected@xdef\@thefnmark{\ref{#1}}\@footnotemark}
\makeatother

\makeatletter \def\url@leostyle{\@ifundefined{selectfont}{\def\UrlFont{\sf}}{\def\UrlFont{\scriptsize\ttfamily}}} \makeatother
\begin{document}

\title{A Cloud Native Platform for Stateful Streaming}

\author{Scott Schneider\textsuperscript{*}}
\author{Xavier Guérin\textsuperscript{*}}
\author{Shaohan Hu}
\author{Kun-Lung Wu}
\affil{IBM T.J. Watson Research Center \\
       Yorktown Heights, NY, USA \\
       \textit{\small \{scott.a.s, klwu\}@us.ibm.com, \{xavier.guerin, shaohan.hu\}@ibm.com}}

\maketitle

\renewcommand*{\thefootnote}{\fnsymbol{footnote}}
\footnotetext[1]{Both authors contributed equally to this work.}
\renewcommand*{\thefootnote}{\arabic{footnote}}

\begin{abstract}

We present the architecture of a cloud native version of IBM Streams, with
Kubernetes as our target platform. Streams is a general purpose streaming system
with its own platform for managing applications and the compute clusters that
execute those applications. Cloud native Streams replaces that platform with
Kubernetes. By using Kubernetes as its platform, Streams is able to offload job
management, life cycle tracking, address translation, fault tolerance and
scheduling. This offloading is possible because we define custom resources that
natively integrate into Kubernetes, allowing Streams to use Kubernetes' eventing
system as its own. We use four design patterns to implement our system:
controllers, conductors, coordinators and causal chains.
Composing controllers, conductors and coordinators allows us to build
deterministic state machines out of an asynchronous distributed system. The
resulting implementation eliminates 75\% of the original platform code. Our
experimental results show that the performance of Kubernetes is an adequate
replacement in most cases, but it has problems with oversubscription, networking
latency, garbage collection and pod recovery.

\end{abstract}

\section{Introduction}

Stream processing enables fast analysis of a high volume of newly arriving data.
While industry and academia have produced many stream processing systems in 
the past two decades~\cite{stream-2003, borealis-2005, spade-2008,
                           millwheel-2013, naiad-2013, storm-2014,
                           heron-2015, flink-2015, dataflow-2015, 
                           flink-2017},
they all share three defining characteristics:

\begin{enumerate}
    \item A programming model that naturally exposes data, task and pipeline
    parallelism. Multiple levels of parallelism allow streaming applications to
    scale with the number of data sources and available hardware.

    \item Interesting streaming applications tend to be stateful. General 
    purpose stream processing can perform non-trivial computations, directly 
    providing answers to users. Such computations typically require
    maintaining state.

    \item A platform and runtime system capable of exploiting the available 
    parallelism while preserving application state.
\end{enumerate}

The kind of platform streaming systems require approaches general purpose
cluster management. Such a platform is responsible for distributing generic user
code across a cluster of compute nodes, arbitrating connections between parts of
the applications on different nodes and managing the life cycle of the
application as well as all component pieces. In the service of managing the
applications on the cluster, streaming platforms must also manage the cluster
itself: track nodes leaving and entering, allocate resources and schedule
applications.

Recently a new kind of technology has emerged for generic cluster management:
cloud platforms built on containers. They are a sweet-spot between clouds that
expose virtual machines and fully hosted single services. Container based cloud
environments are fully generic, but still remove the need for users to manage
an underlying system. Users build container images with programs, libraries and
files in the exact configuration they need.  Users specify how to deploy and
manage the containers that comprise their application through configuration
files understood by the cloud platform.

Because the user is not responsible for managing the cluster of systems in a
container based cloud platform, the cloud platform is. Unlike platforms
for streaming systems, cloud platforms do not \emph{approach} general purpose
cluster management, but \emph{are} general purpose cluster management. One of
the most popular such cloud platforms is Kubernetes~\cite{kube}, which grew out
of Google's experience managing containers for their infrastructure and
applications~\cite{borg-omega-kube-2016}.

Because general purpose cluster management is so valuable, Kubernetes has been
widely adopted by industry to manage their own internal workloads. The
convergence of public cloud platforms with on-premise cluster management is
called \emph{hybrid cloud}. Due to the shared underlying cloud platform,
workloads can migrate between public and private settings.

The ubiquity of cloud platforms like Kubernetes also means there is no escaping
them: they will be used to manage all kinds of software in all settings. For
simple workloads, migrating to cloud platforms simply requires building the
existing software into containers and deploying as needed. Scaling and fault
tolerance are also trivial for simple workloads: just start more containers. But
this approach is not appropriate for workloads such as stream processing that
already have their own management platform and stateful applications. Taking the
simple approach will create a platform-within-a-platform which is difficult to
manage, understand and maintain.

Instead, systems such as streaming platforms need to be rearchitected for cloud 
platforms such as Kubernetes. The new architectures need to be designed around 
the fact that cloud platforms already provide the basics of cluster management.

This paper presents the design of a cloud native version of IBM Streams,
targeting Kubernetes as the cloud platform. Cloud native Streams relies on
Kubernetes for job management, life cycle tracking, scheduling, address
translation and fault tolerance.  The resulting implementation reduces the
platform code base by 75\%. We achieved this reduction in code size by
starting from the question of what existing Streams applications need to
execute, rather than trying to reimplement the existing Streams platform in
Kubernetes. We believe our experience with this rearchitecture will apply to any
system which also required its own management platform. This paper makes the
follow contributions:

\begin{enumerate}
    \item The cloud native patterns used to build cloud native Streams
	(\S~\ref{sec:patterns}). Composing these patterns---controllers,
	conductors, coordinators and causal chains---enable us to construct a
	deterministic platform out of an asynchronous distributed
	system. These patterns are available as an open source library at 
    \url{http://www.github.com/ibm/cloud-native-patterns}.

    \item The architecture of cloud native Streams (\S~\ref{sec:arch}),
    deep-dives on specific features (\S~\ref{sec:features}), and lessons 
    learned (\S~\ref{sec:lessons}).

    \item Experimental results comparing the performance of cloud native
    Streams to a legacy version of Streams (\S~\ref{sec:results}). Network 
    latency, tolerance to oversubscription, garbage collection and pod recovery 
    are worse in the cloud native version. We present these results to help 
    improve Kubernetes.
\end{enumerate}

\section{Related work}

Cloud native applications are defined by their leverage of cloud orchestrators
such as Kubernetes~\cite{kube} or Apache Mesos~\cite{mesos} and their
microservice architecture. Derived from the ``Twelve-Factor App''
guidebook~\cite{12factor}, the microservice architecture is an evolution of the
traditional monolithic and service-oriented architectures, common to enterprise
applications. It favors small, stateless, siloed components with clean
interfaces to maximize horizontal scalability and concurrent development
while minimizing downtimes. The most commonly
published works about cloud-native transformation of legacy workloads cover
stateless applications~\cite{8629101,8457916,8457847,10.1145/3104028,10.1145/3241403.3241440,8004166,KRATZKE20171,7584353,7436659}.

Facebook developed Turbine, which is a cloud native platform for managing their
streaming applications on their Tupperware container
platform~\cite{turbine-2020}.  The main features are a scalable task scheduler,
auto-scaler and consistent and reliable update mechanism. Turbine running on
Tupperware is similar to cloud native Streams running on Kubernetes, with the
exception that Kubernetes handles much more of the platform responsibilities. In
the area of relational databases, {Amazon}~\cite{10.1145/3035918.3056101} and
{Alibaba}~\cite{10.14778/3352063.3352141} undertook the redesign of existing
databases to better fit their respective cloud infrastructure.

For stateful applications, the \emph{lift-and-shift}~\cite{liftshift} approach
is more common than a complete redesign of the supporting platform, often
accompanied with a shim operator that exposes some of the application's concepts
to the cloud platform through the application's native client
interface~\cite{FlinkOperator,BanzaiKafka,StrimziKafka}.

\section{Background and Motivation}

\subsection{IBM Streams}

IBM Streams is a general-purpose, distributed stream processing system. It
allows users to develop, deploy and manage long-running streaming applications
which require high-throughput and low-latency online processing.

The IBM Streams platform grew out of the research work on the Stream Processing
Core~\cite{spc-2006}.  While the platform has changed significantly since then,
that work established the general architecture that Streams still follows today:
job, resource and graph topology management in centralized services; processing
elements (PEs) which contain user code, distributed across all hosts,
communicating over typed input and output ports; brokers publish-subscribe
communication between jobs; and host controllers on each host which
launch PEs on behalf of the platform.

The modern Streams platform approaches general-purpose cluster management, as
shown in Figure~\ref{fig:streams_v4_v6}. The responsibilities of the platform
services include all job and PE life cycle management; domain name resolution
between the PEs; all metrics collection and reporting; host and resource
management; authentication and authorization; and all log collection. The
platform relies on ZooKeeper~\cite{zookeeper} for consistent, durable metadata
storage which it uses for fault tolerance.

Developers write Streams applications in SPL~\cite{spl-2017} which is a
programming language that presents streams, operators and tuples as
abstractions. Operators continuously consume and produce tuples over streams.
SPL allows programmers to write custom logic in their operators, and to invoke
operators from existing toolkits. Compiled SPL applications become archives that
contain: shared libraries for the operators; graph topology metadata which tells
both the platform and the SPL runtime how to connect those operators; and
external dependencies. At runtime, PEs contain one or more operators. Operators
inside of the same PE communicate through function calls or queues. Operators
that run in different PEs communicate over TCP connections that the PEs
establish at startup. PEs learn what operators they contain, and how to connect
to operators in other PEs, at startup from the graph topology metadata provided
by the platform.

We use ``legacy Streams'' to refer to the IBM Streams version 4 family. The
version 5 family is for Kubernetes, but is not cloud native. It uses the
lift-and-shift approach and creates a platform-within-a-platform: it deploys a
containerized version of the legacy Streams platform within Kubernetes.

\subsection{Kubernetes}

Borg~\cite{borg-2015} is a cluster management platform used internally at Google
to schedule, maintain and monitor the applications their internal infrastructure
and external applications depend on. Kubernetes~\cite{kube} is the open-source
successor to Borg that is an industry standard cloud orchestration platform.

From a user's perspective, Kubernetes abstracts running a distributed
application on a cluster of machines. Users package their applications into
containers and deploy those containers to Kubernetes, which runs those
containers in \emph{pods}. Kubernetes handles all life cycle management of pods,
including scheduling, restarting and migration in case of failures.

Internally, Kubernetes tracks all entities as \emph{objects}~\cite{kubeobjects}.
All objects have a name and a specification that describes its desired state.
Kubernetes stores objects in etcd~\cite{etcd}, making them persistent,
highly-available and reliably accessible across the cluster. Objects are exposed
to users through \emph{resources}. All resources can have
\emph{controllers}~\cite{kubecontrollers}, which react to changes in resources.
For example, when a user changes the number of replicas in a
\code{ReplicaSet}, it is the \code{ReplicaSet} controller which makes sure the
desired number of pods are running. Users can extend Kubernetes through
\emph{custom resource definitions} (CRDs)~\cite{kubecrd}. CRDs can contain
arbitrary content, and controllers for a CRD can take any kind of action.

Architecturally, a Kubernetes cluster consists of nodes. Each node runs a
\emph{kubelet} which receives pod creation requests and makes sure that the
requisite containers are running on that node. Nodes also run a
\emph{kube-proxy} which maintains the network rules for that node on behalf of
the pods. The \emph{kube-api-server} is the central point of contact: it
receives API requests, stores objects in etcd, asks the scheduler to schedule
pods, and talks to the kubelets and kube-proxies on each node. Finally,
\emph{namespaces} logically partition the cluster. Objects which should not know
about each other live in separate namespaces, which allows them to share the
same physical infrastructure without interference.

\subsection{Motivation}
\label{sec:motivation}

Systems like Kubernetes are commonly called ``container orchestration''
platforms. We find that characterization reductive to the point of being
misleading; no one would describe operating systems as ``binary executable
orchestration.'' We adopt the idea from Verma et al.~\cite{borg-2015} that
systems like Kubernetes are ``the kernel of a distributed system.'' Through CRDs
and their controllers, Kubernetes provides state-as-a-service in a distributed
system. Architectures like the one we propose are the result of taking that view 
seriously.

The Streams legacy platform has obvious parallels to the Kubernetes
architecture, and that is not a coincidence: they solve similar problems.
Both are designed to abstract running arbitrary user-code across a distributed
system.  We suspect that Streams is not unique, and that there are many
non-trivial platforms which have to provide similar levels of cluster
management.  The benefits to being cloud native and offloading the platform
to an existing cloud management system are: 
\begin{itemize}
    \item Significantly less platform code.
    \item Better scheduling and resource management, as all services on the cluster are 
        scheduled by one platform.
    \item Easier service integration.
    \item Standardized management, logging and metrics.
\end{itemize}
The rest of this paper presents the design of replacing the legacy Streams 
platform with Kubernetes itself.

\section{Cloud Native Patterns}
\label{sec:patterns}

We present the key abstractions that made the migration of IBM Streams from a
legacy platform to a cloud native platform possible: controllers, conductors,
coordinators and causal chains. The patterns are available as an open source
Java library at \url{http://www.github.com/ibm/cloud-native-patterns}.

\subsection{Controllers}
\label{sec:controller}

Kubernetes defines controllers as ``control loops that tracks at least one
resource type''~\cite{kubecontrollers}. We constrain that definition further: in
cloud native Streams, a controller is a control loop that tracks a single
resource type. Controllers take some action on creation, modification and
deletion of a resource type. As with regular resources, custom resources can be
monitored using controllers. Cloud native Streams makes extensive use of custom
resources to store platform related state (Figure~\ref{fig:actors}).
For each custom resource, it implements a concurrent controller deployed within
the instance operator. The instance operator also uses controllers
for traditional Kubernetes resources such as pods, nodes, and namespaces.

In cloud native Streams, a resource controller is responsible for monitoring
events on a single resource, saving state updates for this resource in a local
cache, and executing any action as a result of these events. We use the
microBean Kubernetes Controller library~\cite{microbean} as a starting point for
our controller abstraction. It implements both the informer and reflector
patterns as defined in the Go client~\cite{kubegoclient}.  In its most essential
form, a cloud native Streams controller is built by defining a set of three
event callbacks (\code{onAddition}, \code{onModification}, \code{onDeletion}), a
resource store, and providing both callbacks and store to an internal microBean
controller instance.

\begin{figure}[ht!]
  \centering
  \includegraphics[width=1.0\linewidth]{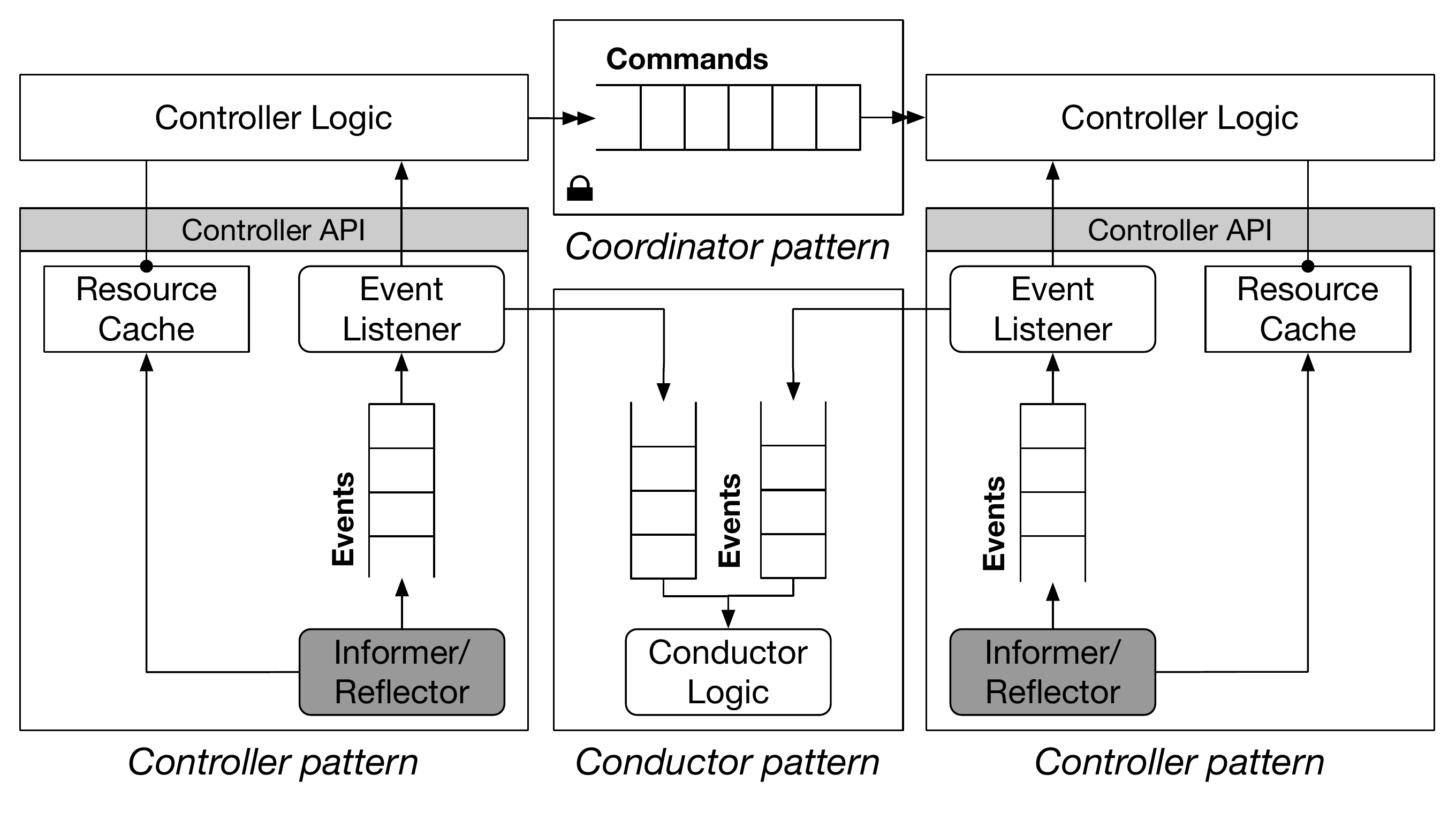}
  \caption{Clound native patterns: controller, conductor and coordinator. This example shows 
    two controllers and one conductor. The controller on the left uses the coordinator pattern to 
    update resources owned by the controller on the right. A conductor monitors events from both 
    resources.}
  \label{fig:patterns}
\end{figure}

The \emph{controller pattern} in Figure~\ref{fig:patterns} depicts the
relationships between these components. An \emph{event listener}
categorizes notifications it receives from the microBean informer into addition,
modification and deletion events on resources. Cloud native Streams
controllers, which derives from an event listener, implement the actions
to take in response to these events. The resource cache is used by the microBean
reflector to maintain a cached view of the resource pool based on the streams of
event it has received.

\subsection{Conductors}
\label{sec:conductor}

In contrast to controllers, the \emph{conductor pattern} (bottom-middle of
Figure~\ref{fig:patterns}) observes events from multiple resources and does not
save state updates in a local cache. Instead, they are concurrent control loops
that maintain a state machine that transitions based on resource events, all
towards a final goal. Conductors do not own any resources. Rather, they register
themselves with existing controllers as generic event listeners which receive
the same notifications that each controller does.

The conductor pattern solves the problem of synchronizing a particular action
based on asynchronous events generated by multiple actors. In cloud native
Streams, we encountered the conductor pattern in two main cases: job submission
and pod creation for PEs. For jobs, we need to know when to move the job's
status from the initial \code{Submitting} to the final \code{Submitted} state.
A job is not fully submitted until all of the resources that comprise it have
been fully created. Before we can create a pod for a PE, we must first ensure
that all of its dependencies exist, such as secrets or its \code{ConfigMap}.

In both cases, it is necessary to listen to multiple resource events and
maintain local tracking of the status of those resources in order to arrive at
the goal.

\subsection{Coordinators}
\label{sec:coordinator}

When asynchronous agents need to modify the same resource, we use the
\emph{coordinator pattern} (top-middle of Figure~\ref{fig:patterns}). The
coordinator pattern implements a multiple-reader, single-writer access model by
granting ownership of the resource to a single agent and serializing
asynchronous modification requests coming from other agents. Coordinators are
synchronous command queues that serially execute modification commands on
resources. In cloud native Streams, this pattern means that the controller for a
resource owns that resource, and other controllers which want to modify it must
make requests to that controller.

Many situations in cloud native Streams involve concurrent agents wanting to
modify resource they interact with. For example, a PE's launch count tracks how 
often the platform has started a PE. There are two instances when we must increment 
the PE launch count: pod failure and PE deletion. However, different agents handle 
those two events. Allowing them to asycnhronsouly modify the PE's launch would 
lead to race conditions.

Instead, we use the coordinator pattern. When a pod fails, it must be restarted.
Our pod controller overrides the default behavior of letting the kubelet restart
the pod. When the pod controller is notified of a failed pod, it must increase the restart
count of its owning PE. Doing so directly through a Kubernetes update command
could lead to race conditions between the PE controller and the pod controller.
Instead, we use the PE coordinator interface to let the PE controller execute
that command for us.

\subsection{Causal Chains}
\label{sec:causal}

\begin{figure}[h]
  \centering
  \includegraphics[width=1\linewidth]{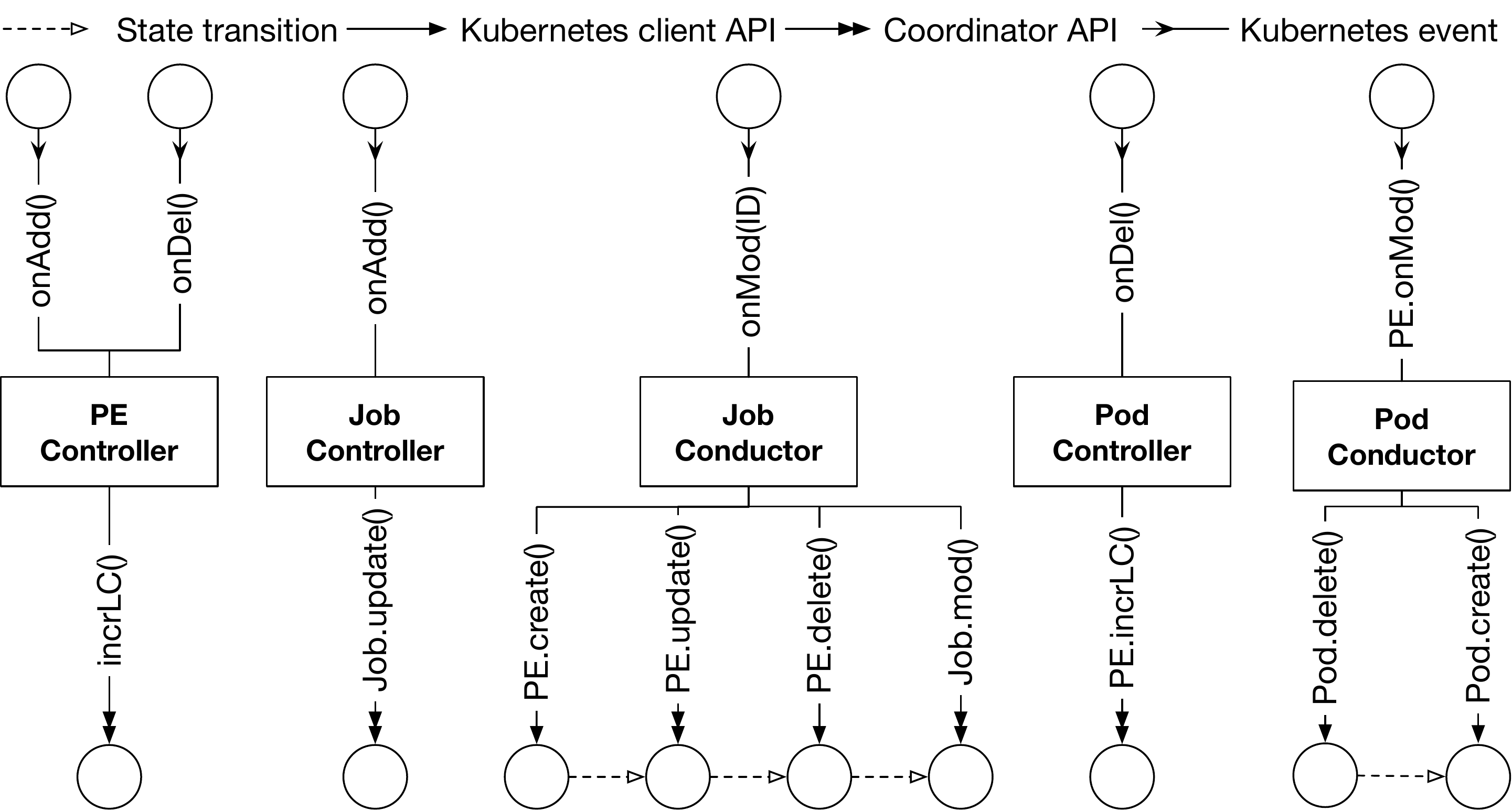}
  \caption{Causal Links}
  \label{fig:causal_links}
\end{figure}

\begin{figure}[h]
  \centering
  \includegraphics[width=1\linewidth]{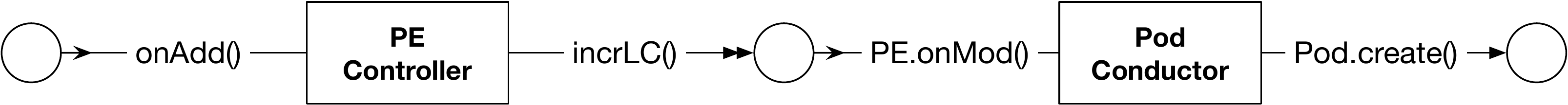}
  \caption{Causal Chain}
  \label{fig:causal_chain}
\end{figure}

A \emph{causal link} (Figure~\ref{fig:causal_links}) is a single actor
responding to a single resource change by synchronously changing one or more
other resources. These logical state transitions are atomic and composable. A
\emph{causal chain} (Figure~\ref{fig:causal_chain}) is the composition of
multiple causal links where the result of one is the input to another. It is an
asynchronous sequence of deterministic actions that implements a logical state transition
across multiple resources. Unlike the other patterns, causal links and chain are
not themselves actors in the system.  They are a pattern of behavior that
emerges from the interaction of multiple actors.

Causal chains are an abstraction that derives from two principles: operational
states have a single source of truth provided by Kubernetes, and these states
can only be synchronously modified in a single place in the system. As a consequence, 
causal chains necessarily span across multiple actors: modifications initiated by one 
actor on a resource it controls cause another actor to make modifications to its 
own resource.

\begin{figure*}[t!]
  \centering
  \includegraphics[width=0.7\linewidth]{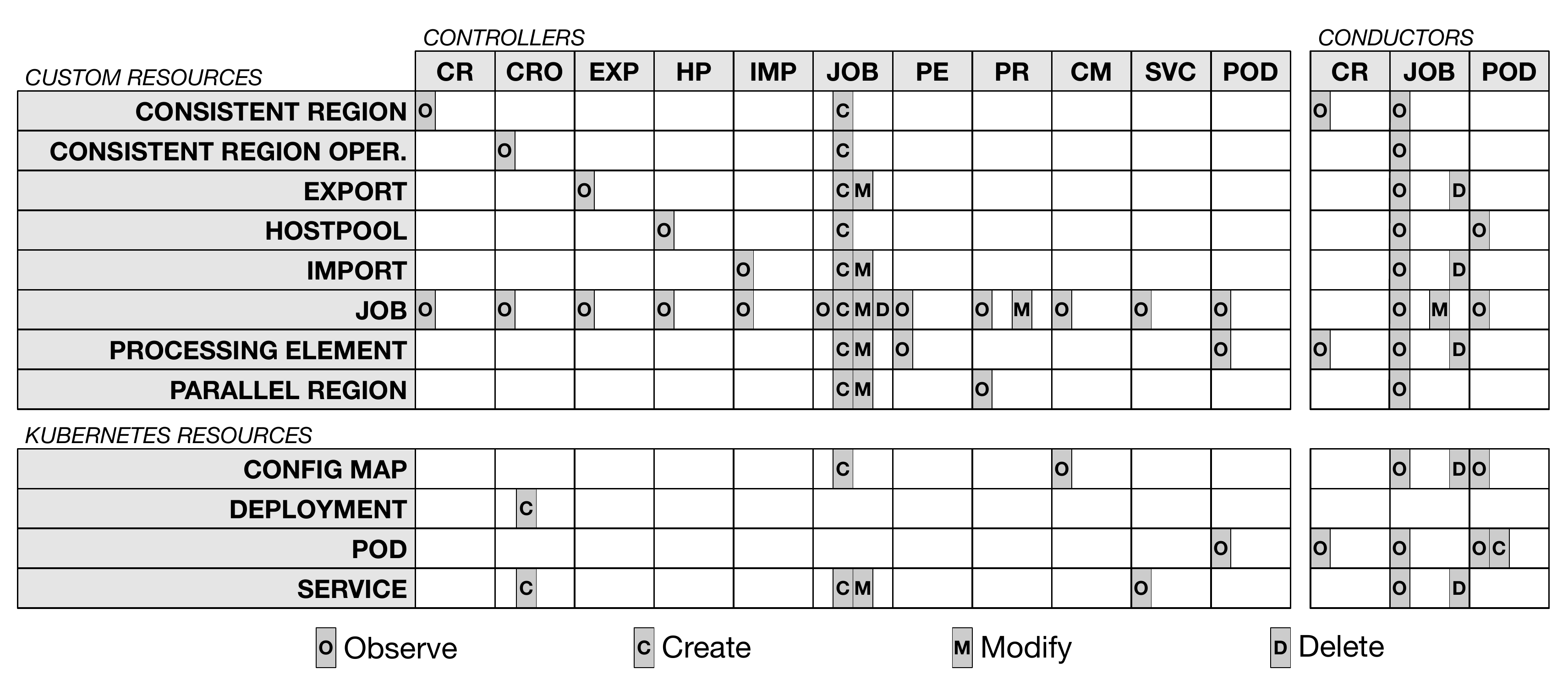}
  \caption{Clound native Streams actors and their interactions.}
  \label{fig:actors}
\end{figure*}

In cloud native Streams, a causal link drives the creation of a pod for a PE.
The pod conductor is the only actor which can create pods for PEs, and it only 
reacts to changes to a PE's launch count. It composes with four other causal 
links to create causal chains:

\begin{enumerate}
    \item \emph{PE creation}. The PE controller reacts to a new PE by incrementing 
        the launch count through the PE coordinator.
    \item \emph{Voluntarily PE deletion}. The PE controller recreates the PE resource, 
        which eventually leads to (1).
    \item \emph{Pod failure or deletion}. The pod controller increments the PE 
        launch count through the PE coordinator.
    \item \emph{Job submission}. The job conductor checks if any PEs for the job are 
        already running, and if they are, if the graph metadata for them has 
        changed. If yes, then the job conductor updates the graph metadata and 
        increments the launch count through the PE coordinator.
\end{enumerate}

By composing controllers, conductors and coordinators, we construct a
deterministic state machine out of an asynchronous distributed system.
Controllers are state machines that react to changes on a single resource kind,
but may produce changes on any resource. Conductors are state machines that
react to and produce changes on multiple resource kinds.  The composition of
controllers and conductors is necessarily itself a state machine, but it is the
addition of coordinators that makes the resulting state machine deterministic.

\section{Architecture}
\label{sec:arch}
Four goals guided our design:

\begin{enumerate}
    \item \emph{Discoverability}. Users should be able to use their pre-existing 
        knowledge of Kubernetes to discover what their application is doing, how 
        to modify it, and how Streams works in general. Streams applications 
        should also be discoverable by other workloads on the same Kubernetes 
        cluster, through standard Kubernetes mechanisms.
    \item \emph{Composability}. By using Kubernetes first-class service
        endpoints, cloud native Streams should interoperate with other applications
        and middleware without further configuration. 
    \item \emph{Application state preservation}. Stateless services are easy to
        manage in Kubernetes, as simply restarting them is always an option.
        But most Streams applications have state. We have an implicit contract
        with users that once they deploy an application, they will not lose any
        accumulated state---barring application failure and users taking
        explicit action to restart it.
    \item \emph{Backwards compatibility}. The cloud native version of Streams 
        should run legacy Streams applications unchanged. This goal means that 
        we cannot change any public APIs, and our task is to find the most 
        Kubernetes-like way to express functionality originally designed 
        for an on-premises cluster.
\end{enumerate}

We also have one anti-goal: we do not maintain API compatibility with the legacy
platform. Trying to do so would force cloud native Streams to take on
responsibilities that should belong to Kubernetes, which would end up
conflicting with our stated goals. We want user's \emph{applications} to remain
unchanged, but we are assuming that they are adopting cloud native Streams as
part of an overall effort to consolidate and simplify management and
administration.

\subsection{Overview}

All aspects of a Streams application exposed to users are represented as CRDs or
existing Kubernetes resources. We apply the patterns described in
\S~\ref{sec:patterns}: each resource is managed by a controller; when we
need to monitor the status of multiple kinds of resources in order to take an
action we use a conductor; and when multiple actors need to change the state for
a particular resource, we use a coordinator.

The CRD is the foundational unit in our design. CRDs are exposed to users in the
same way as any other Kubernetes resource, which means that representing Streams
concepts as CRDs gains not just native integration into the Kubernetes system,
but also the user interfaces. Any state that we must maintain goes into a CRD;
all state not in CRDs is ephemeral and can be lost without consequence.
Kubernetes delivers reliable event notifications when CRDs and other resources
are created, deleted and modified. Reacting to these notifications in
controllers and conductors is the primary communication mechanism between all of
the actors in our system.

The CRDs in Figure~\ref{fig:actors} define the following resources:

\begin{itemize}
    \item \code{Job}: A single Streams job. The job 
        controller initiates the job submission and tracks unique job identifiers. 
        The job conductor manages the job submission process and update 
        its status when completed.
    \item \code{ProcessingElement}: A PE in a job. The 
        PE controller tracks launch count and restores voluntarily deleted PEs.
    \item \code{ParallelRegion}: A single parallel region in a job. It 
        exposes a \code{width} attribute that can be directly altered by users using 
        \code{kubectl edit} or the Kubernetes client API. The parallel region
        controller handles width changes applied to parallel regions.
   \item \code{HostPool}: A host pool in a job.
   \item \code{Import}: An imported stream in a job. The import
       controller monitors the addition and modification of these resources and 
       matches them with export resources.
   \item \code{Export}: An exported stream in a job. The export
       controller monitors the addition and modification of these resources and 
       matches them with import resources.
    \item \code{ConsistentRegion}: A consistent region in a job. The 
        consistent region controller coordinates application checkpoints and 
        restarts for a single region.
    \item \code{ConsistentRegionOperator}: tracks all consistent regions in a 
        job. Created on-demand during a job submission with a consistent region.
        Its controllers monitor the deployments used to create the operators.
\end{itemize}

We also leverage the following Kubernetes resources:

\begin{itemize}
    \item \code{ConfigMap}: Shares job specific configuration between controllers and pods, 
        such as the graph metadata used by PEs to inform them of the operators they contain 
        and how to connect to other PEs. 
    \item \code{Deployment}: Manages the instance operator and the consistent region operator.
    \item \code{Pod}: Executes PEs. We use a pod controller to monitor and 
        manage the life cycle of pods within jobs. The pod conductor waits until all required
        resources are available before starting a pod for a PE.
    \item \code{Service}: Exports PE entrypoints as well as user-defined 
        services within PEs.
\end{itemize}

Figure~\ref{fig:actors} shows how our actors interact with each other. There
are four kinds of actions they can take:

\begin{enumerate}
    \item \emph{observes}: the actor either receives events from Kubernetes about that 
        resource, or passively views its store.
    \item \emph{creates}: the actor creates new instances of the resource through 
        commands to Kubernetes.
    \item \emph{deletes}: the actor deletes particular instances of a resource 
        through commands to Kubernetes.
    \item \emph{modifies}: the actor makes changes to an already created resource 
        through that resource's coordinator.
\end{enumerate}

None of our actors communicate directly with each other; all communication happens 
by creating, modifying or deleting Kubernetes resources.

\begin{figure}[t!]
  \centering
  \includegraphics[width=1\linewidth]{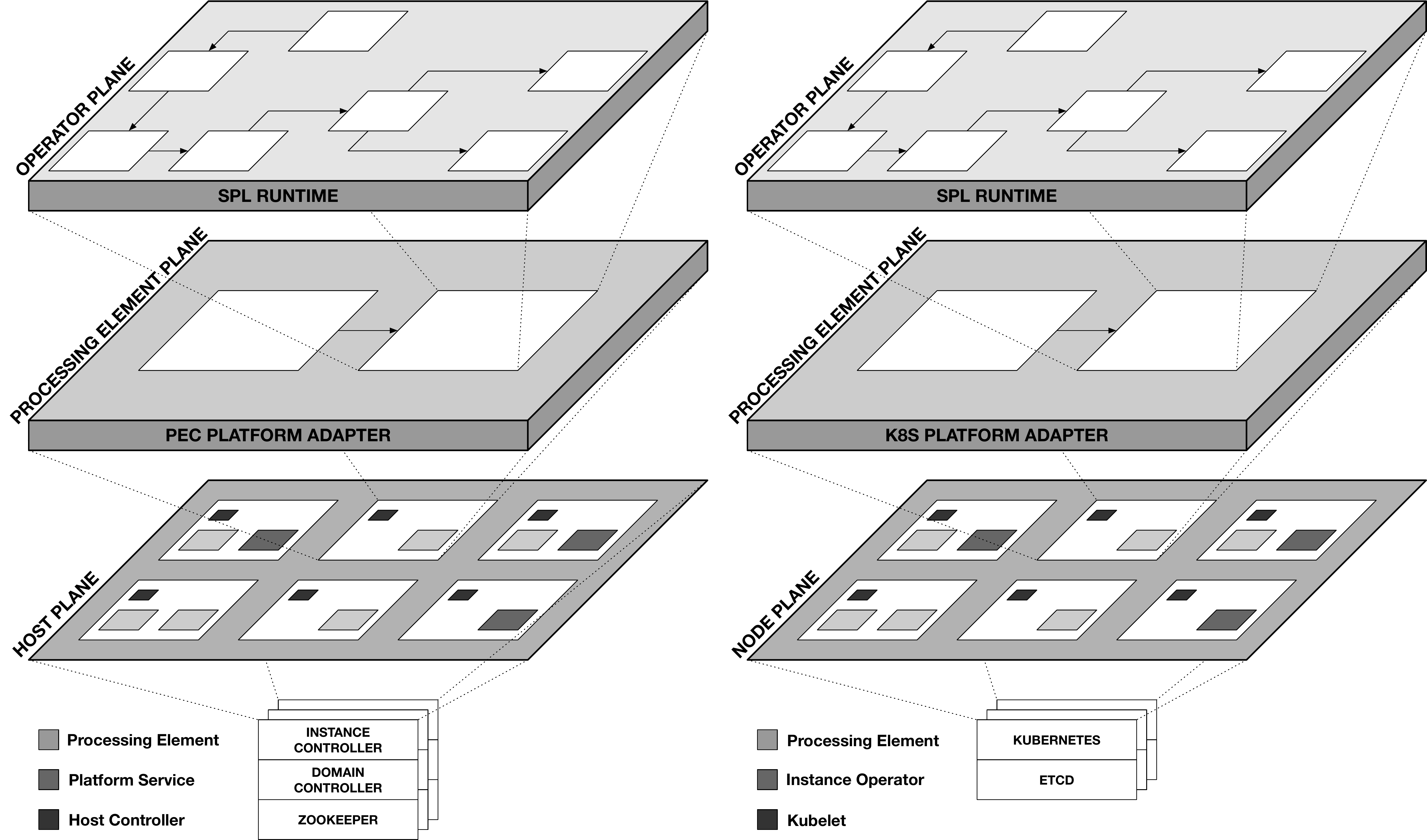}
  \caption{Legacy and cloud native Streams deployments}
  \label{fig:streams_v4_v6}
\end{figure}

Figure~\ref{fig:streams_v4_v6} depicts the deployed artifacts of an instance of cloud
native Streams.  The Streams \emph{instance operator} contains all the
controllers, conductors and coordinators for a Streams instance. Each Kubernetes
namespace can have one Streams instance operator. The instance operator maps to
the legacy concept of an instance. The legacy concept of a Streams
domain---management of Streams instances---is no longer needed as the Kubernetes
cluster serves that role.

In Streams, the PE is the vehicle for executing user code. PEs contain an
arbitrary number of user operators and the application runtime.  In our design,
we always assign one PE to a pod. This design decision is fundamental. It
allows us to: tie a runtime PE's life cycle to that of the pod that contains
it; fully offload PE scheduling to Kubernetes; rely on the Kubernetes DNS
service for establishing direct TCP connections between our PEs.  Any other
design would have required bespoke implementations of life cycle management,
scheduling and network name resolution.

While the PE runtime is unchanged, we had to implement a new translation layer
between the PE and Kubernetes. It implements the platform abstraction for the PE
runtime, as well as instantiates and initializes the PE. At runtime, it
also: collects metrics from the PE using a pre-existing interface and exposes
them to Prometheus~\cite{prometheus}; monitors and reports the status of all
PE-to-PE connections; and monitors liveness and reports it to Kubernetes.

\subsection{Loose Coupling}
\label{sec:coupling}

The legacy Streams platform was tightly coupled, which lead to operational
difficulty and implementation complexity. Cloud native Streams applies the
concept of loose coupling.

\paragraph{Name resolution:} PEs communicate with each other over TCP
connections. It is the platform's responsibility to define PE ports, give them
names, and allow PEs to find each other's ports by those names.

In legacy Streams, each PE port is assigned a \code{(peId, portId)} tuple called
a \emph{port label} that uniquely identifies that port in the instance.
At initialization, PEs must establish their remote connections to other PEs
using their port labels. To that end, each PE first creates the socket receiver
for each of its receiver ports, determines its local TCP port, and publishes to 
ZooKeeper its mapping of port label to \code{(hostname, tcpPort)}. PEs already
know the port label that each sending port needs to connect to through graph
metadata that the platform provides at PE startup. After publishing its own
receiver port labels, each PE then looks up the translation of the remote
receiver port label in ZooKeeper for each of its own sender ports and
establishes those connections.  Even with some caching (used to reestablish lost
connections), the thundering herd aspect of this initialization process and the
strain it applies to the ZooKeeper ensemble delays initial deployment times.

Cloud native Streams relies on the Kubernetes name resolution system to resolve
inter-PE connections. There are similarities with the legacy system name
resolution system, as it relies on etcd to store its Service configurations, and
it also has some currently unresolved latency issues~\cite{codacydns,machudns}.
But from an application perspective, it is easier to use as name resolution is
done using standard BSD functions such as \code{gethostbyname()}. Lastly, from
an administration perspective, it is simpler to manage as it binds directly to
the container's \code{/etc/resolv.conf} subsystem and can be easily superseded.

\paragraph{Message bus:} The message bus in legacy Streams between the platform
and PEs uses full-duplex, synchronous communication channels implemented with
JMX~\cite{jmx}. All initiated communications must succeed. Failed
communications are retried with increasing backoff delays before being
escalated as more general system failures where it may restart a PE. As job
count and PE size increases, communications tend to time out more frequently,
leading to failure escalations reducing responsiveness. We have witnessed tens
of minutes to list all the PEs in an overloaded instance.

Cloud native Streams decouples the instance operator from the PEs by relying on
the states stored in Kubernetes to achieve operational availability. The
controller pattern (\S~\ref{sec:controller}) is used by all agents
interested in keeping track of those states. Agents that need to notify the
instance operator of internal state changes do so through Kubernetes events.  In
turn, the instance operator synchronously applies those changes to the custom
resources, preventing potential race conditions (\S~\ref{sec:coordinator}).

The SPL runtime, including the PEs, are implemented in C++. As of this writing,
no library capable of implementing our controller pattern is available in C/C++.
As an alternative, we temporarily resort to a set of REST services hosted by the
instance operator. State changes within agents other than the instance operator
periodically send REST operations to those services to notify the operator of
internal changes. In turn, the operator applies these changes to the related
custom resources. Implementing the proper library and removing the REST layer is
part of our future work.

\subsection{Fault Tolerance and Rolling Upgrades}

Fault tolerance and general high-availability is a primary goal in the design of
Streams since streaming applications are expected to run for months without
interruption. To that end, the legacy Streams platform was designed such that:

\begin{enumerate}
    \item All platform related state is persisted in ZooKeeper. Upon failure, 
        platform services restart and retrieve their state from ZooKeeper.
    \item Streaming applications continue to run during platform service
        failures or upgrades.
    \item Applications seamlessly resume operations after the loss of a
        PE or a host.
\end{enumerate}

In cloud native Streams, we use Kubernetes to preserve
these attributes.

\paragraph{Persistent states:} Kubernetes exposes state persistence to users
through CRDs. Cloud native Streams makes heavy use of CRDs 
to maintain states critical for recovery. However, where the legacy
platform implementation favored storing the state of the system as-is,
cloud native Streams stores only what is necessary and sufficient
to reach the current state of the system through recomputation. The reasons
behind that radical shift in the computation versus space trade-off of
our system are:

\begin{enumerate}
    \item We discovered through empirical measurements that the
        amount of time required to perform state recomputation is negligible
        compared to other operations in the system and appear instantaneous to
        human users.
    \item Minimizing the amount of data persisted drastically reduces the
        pressure on the persistent ensemble.
    \item Re-computing intermediate state simplifies the design of our system.
\end{enumerate}

\paragraph{Instance operator:} The Streams instance operator is designed to be
resilient to its pod restarting.  All of the actors in the instance operator
will receive the full history of Kubernetes events that they are subscribed to,
allowing them to catch-up to the current state of the system.  The applications
themselves do not need the instance operator for normal operation, so they can
continue unharmed. Because of this resiliency, the instance operator can easily
recover from failure. Upgrades are also trivial: change the image for the
instance operator and restart the pod. The combination of how we defined our
CRDs, the patterns we use to manage them and Kubernetes' reliable event delivery
enable these capabilities.

\paragraph{Applications:} We consider two types of application failures:
voluntary failures, when a user deletes a resource; and involuntary failures,
when a PE crashes or a node becomes unavailable. The voluntary deletion of job
resources are caught by the \code{onDeletion()} callback in that resource's
controller. In this situation, the deleted resource is recreated
by the controller if the owning job exists and is in the \code{Submitted} state.

Special care needs to be taken in the event of pod failure or PE deletion. To
maintain Streams' application consistency logic
(\S~\ref{sec:consistentregion}), restarting a pod needs to be coordinated with
both the PE and pod controllers through a causal chain (\S~\ref{sec:causal}). 

\section{Feature Deep-Dives}
\label{sec:features}

\begin{figure*}[t!]
    \centering
    \includegraphics[width=0.75\linewidth]{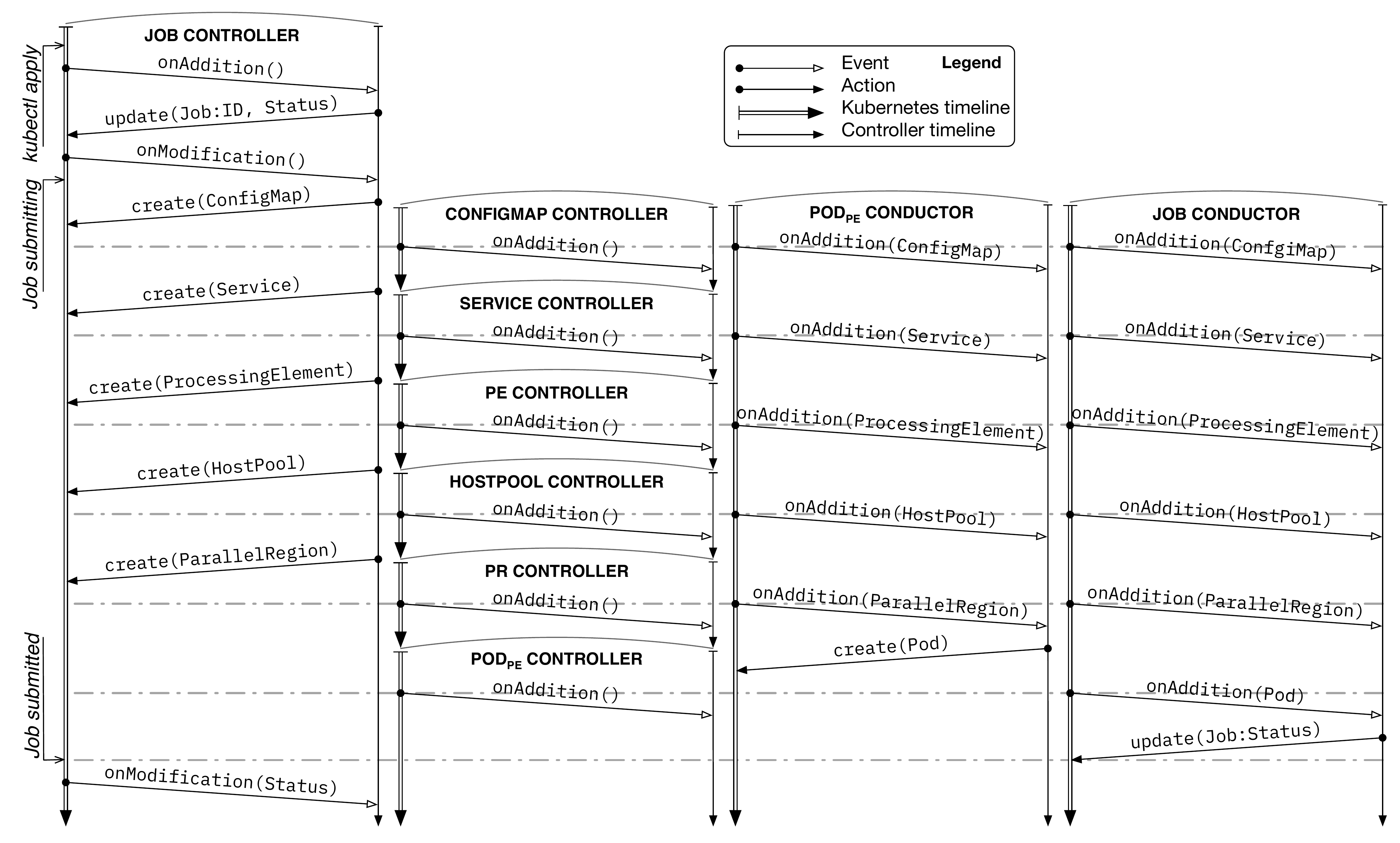}
    \caption{Job submission event diagram.}
    \label{fig:submission}
\end{figure*}

\subsection{Job Submission}
\label{sec:jobsubmission}

Users submit compiled application archives to create running jobs.  During job
submission, the platform must:

\begin{enumerate}
    \item Create an internal logical model of the application by extracting the 
        graph metadata from the application archive. This logical model is a 
        graph where each node is an operator and all edges are streams. Some 
        operators in the logical graph will not execute as literal operators, 
        as they represent features of the application runtime.
    \item Transform the logical model based on application features, such as 
        parallelism or consistent regions.
    \item Generate a topology model from the logical model. The topology model is 
        a graph where all of the nodes are executable operators. This process
        requires turning logical operators that represent application runtime features 
        into metadata on executable operators.
    \item Fuse the topology model into PEs. Each PE is an independent 
        schedulable unit that contains at least one operator. Fusion also 
        requires creating PE ports: streams between operators that cross PE 
        boundaries require unique ports. At runtime, these ports will send and 
        receive data across the network.
    \item Generate graph metadata for each PE that tells it at runtime how 
        to connect operators inside and outside of it.
    \item Schedule and place the PEs across the cluster.
    \item Track job submission progress.
\end{enumerate}

\paragraph{Legacy:} Users submitted jobs to the legacy platform through a
bespoke command line tool or a development environment that communicated with
the platform over JMX. This process was synchronous and monolithic. The entire
process would not return until the job was either scheduled and placed, or some
failure prevented that.

During job submission, PEs were given IDs that were globally unique in the
instance. A PE's ports were assigned IDs that were unique across that job.
Upon creation, the topology model was immediately stored in ZooKeeper. The
manner of its storage was fine-grained: each node and edge was individually
stored in ZooKeeper. A job's topology model, backed by ZooKeeper, was kept
around and actively maintained for the lifetime of the job.

\paragraph{Cloud native:} Since jobs are represented as CRDs, users submit new
jobs through the \code{kubectl apply} command, or programmatically through any
Kubernetes client.

Figure~\ref{fig:submission} is a job submission event diagram. Note that none of
the controllers or conductors talk directly to each other: they exclusively
interact with Kubernetes by creating, modifying or viewing resources. Kubernetes
then delivers events based on these resources to all listening actors. When the
job controller receives a new job notification, it executes steps 1--5. The code
for executing these steps is reused from the legacy platform, with three major
changes: PE IDs are not globally unique in the instance, but are local to the
job; PE port IDs are not unique within the job but are local to the PE; and the
job topology is not stored anywhere. In fact, the job topology is
discarded once the job controller extracts the metadata and stores it in a local
context.  Finally, the job controller assigns a job ID and marks it as
\code{Submitting} by updating the job CRD.

The job controller does not create any resources for the job until it has a
guarantee that Kubernetes has successfully stored the new ID and job status.  It
ensures this is true by waiting for the modification notification from
Kubernetes.  It then uses the job metadata in the local context to start
creating resources. Note that this local context truly is local and ephemeral.
If the job controller fails before a job is fully submitted, upon restart, it
will delete the partially created resources, create a copy of the job CRD,
delete the original, and create a new job through Kubernetes.  It then goes
through the submission process again. Rather than trying to save progress along
the way, it is simpler to lose and delete transitory state and then restart the
process over again.

That all actors work asynchronously is evident Figure~\ref{fig:submission}'s
timeline.  But some actions need to happen synchronously, such as pod creation:
we can only create pods for PEs when we know that all of the other resources
that pod depends on already exist.  The conductor pattern solves this problem.
The pod conductor receives creation events for all of the resources that a PE's
pod needs, and only when all of those resources exist does the pod conductor
create the pod for that PE.

The job conductor solves a similar problem for job status.  The job submission
process must report its status to users.  But status is also important
internally: because of the stateful nature of Streams applications, once a job
has successfully submitted, simply deleting resources and starting over is no
longer a viable method for dealing with updates or failures. The job conductor
tracks the creation status of all resources that comprise a job, and when all
exist, it commits the job to the \code{Submitted} state by updating the CRD with
Kubernetes.

\subsection{Scheduling}

SPL allows users to control PE scheduling~\cite{spl_placement}: PEs can
be colocated, exlocated, run in isolation or assigned to a \emph{hostpool},
which is an SPL abstraction for a set of hosts. The platform is responsible for
honoring these constraints while also scheduling the PEs across the cluster in a
balanced manner.

\paragraph{Legacy:} Since legacy Streams assumes that it owns the cluster, it
was responsible for scheduling each PE on a host. The scheduler performs a
finite number of attempts to find suitable hosts for each PE. Each round uses a
different heuristic for how to favor PE placement. It tries to find a suitable
host for each PE while honoring the constraints for that PE, and the constraints
of the PEs already placed on hosts. The default behavior of the scheduler is to
balance PEs proportional to the number of logical cores on a host while
considering the PEs already placed from previous jobs.

\paragraph{Cloud native:} As every PE is in its own pod, Kubernetes handles
scheduling. Our responsibility is to communicate the PE constraints originally
specified in the SPL application to the Kubernetes pod scheduler. The
natural solution is through a pod's spec. We map the following existing SPL
scheduling semantics onto the mechanisms exposed by pod specs:

\emph{Host assignment:} We map the concept of a physical host in legacy Streams
to a Kubernetes node in cloud native Streams. For PEs that request specific
node names, we use the \code{nodeName} field in the pod spec.  This mapping is
natural, but requires justification: in an ideal cloud native environment,
users should not need to care about what nodes their code runs on. But a
use-case for legacy Streams still applies in a cloud environment: specific
nodes may have special capabilities such as hardware acceleration that PEs
require.

SPL also has the concept of a tagged hostpool: PEs do not request a specific
host, but rather any from a set of hosts with a specific tag. The concept of
tags maps directly to Kubernetes labels, which we can use with the
\code{nodeAffinity} option in a pod's \code{affinity} spec.


\emph{Colocation:} PEs request colocation with other PEs through using a
common token. They don't care about what host they run on, as long as they are
scheduled with other PEs that specify that token. We can achieve the same
scheduling semantics using pod labels and \code{podAffinity} in the pod spec:
generate a unique Kubernetes label for each token in the application, and
specify that label in \code{podAffinity} in the pod spec. Together, both halves
implement the full semantics: \code{podAffinity} ensures that this PE
is scheduled on the same node as PEs with the same label, and the label
ensures that all other PEs with matching affinity are scheduled with this PE.

\emph{Exlocation:} PEs request exlocation from other PEs through a common token.
All PEs which exlocate using the same token will run on different hosts. We
achieve these semantics in Kubernetes by using the same scheme as with
colocation, except we use \code{podAntiAffinity} in the pod spec.

\emph{Isolation:} PEs can request isolation from other PEs, but pod specs do not
have a single equivalent mechanism. However, note that requesting isolation from
all other PEs is semantically equivalent to requesting exlocation from each PE
individually, using a unique token for each pairing. We further note that
exlocation is symmetric and transitive. It is symmetric because if two PEs are
exlocated from each other, they both must have requested exlocation with the
same token. And it is transitive because if \emph{A} is exlocated from \emph{B},
and \emph{B} is exlocated from \emph{C}, then \emph{A} must be exlocated from
\emph{C}. However, the \code{podAntiAffinity} spec is not symmetric: if pod
\emph{A} specifies anti-affinity to \emph{B}, that does not require \emph{B} to
specify anti-affinity to \emph{A}. Because the pod relationship is not
symmetric, we avoid transitivity. From this insight we can build PE isolation
through pod labels and \code{podAntiAffinity}.  For each isolation request in a
job, create a unique label. We apply this label to each PE's pod spec, except
for the PE that requested isolation. For the requesting PE, we use
\code{podAntiAffinity} against that label.

%

\subsection{Parallel Region Updates}
\label{sec:parupdates}

SPL allows developers to annotate portions of their stream graph as
\emph{parallel regions}~\cite{spl_parallel}. Parallel expansion during job
submission replicates all of the operators in such regions, and the
runtime partitions tuples to different replicas to improve tuple processing
throughput through data parallelism.

Users can dynamically change the width of a parallel region, growing or
shrinking the number of replicas. The platform will restart all PEs with
operators in the parallel region, and all PEs with operators that communicate
directly with them. (The PE runtime cannot dynamically its stream graph, so we
must restart them to apply changes.) However, all other operators in the
application should stay up. If we did not need to keep the operators outside of
the parallel region running, we could trivially achieve a parallel region
update by re-submitting the job with the new width and restarting everything.
The process to find which operators to add, remove and modify is:

\begin{enumerate}
    \item Re-generate the logical and topology model of the application with the 
        original parallel width.
    \item Generate the logical and topology model of the application with the 
        new parallel width.
    \item Perform a diff of the selected parallel region across both topology models, 
        figuring out which operators were added, removed or changed.
    \item Graft the target parallel region from the topology model with the new width
        into the original model.
    \item Re-index all of the operators and streams in the parallel region as 
        necessary to maintain consistency with the original topology model.
\end{enumerate}

After determining the affected operators, the platform is responsible for
figuring out how to add, remove or restart the PEs with them.

\paragraph{Legacy:} Users changed the parallel width for a region in a running
job through either a command line tool or a development environment connected
with the platform over JMX.

The legacy Streams platform was not designed for dynamic job topology changes.
But two key design details made it particularly difficult: PE IDs are unique
within the instance, and PE port IDs are unique within the job. As a result,
dynamic changes cannot go through the same code path as job submissions.  Trying
to do so would result in assigning new IDs to unchanged PEs and ports, which
would require restarting them. Instead, the legacy platform goes through a
separate process for dynamic updates where only the changed operators are
considered for fusion, scheduling and placement.

\paragraph{Cloud native:} Kubernetes was designed for dynamic updates; updating
a resource is a standard operation. We take advantage of this design because we
represent parallel regions as CRDs. Users can edit the parallel region CRD's
width through \code{kubectl} or a Kubernetes client. The parallel region
controller will then receive the modification notification.

In cloud native Streams, PE IDs are local to the job, and PE port IDs are local
to the PE. For example, if a Streams job has two PEs, their IDs are always 0 and
1. If a PE has a single input port, its ID is always 0. If a PE has $n$ output
ports, an additional output port will always be output port $n$. This
deterministic naming also means that it is necessarily hierarchical: in order to
refer to a particular PE, we must also refer to its job, and in order to refer
to a particular port, we must also refer to its PE.  We store this graph
metadata in the \code{ConfigMap} for each PE's pod, and at runtime, PEs use this
graph metadata to establish connections between each other.

This seemingly minor design point allows us to greatly simplify parallel region
updates: the parallel region controller simply feeds the topology model from
step 5 into the normal job submission process through the job coordinator.  Our
job submission process is generation-aware: each generation gets a
monotonically increasing generation ID. We also do not blindly create
resources, but instead use the create-or-replace model where if we try to
create a resource that already exists, we instead modify it.  When the parallel
region controller initiates a new generation for a job, the \code{ConfigMap}s
for the PEs which should not be restarted will have identical graph metadata as
before, due to our deterministic naming scheme.  The pod conductor remains
active, even after a successful job submission. It will receive modification
notifications for these \code{ConfigMap}s for each PE.  If the graph metadata
is identical to the previous generation, it will update the generation ID for
the pod, and take no further action. If the graph metadata is different, it
will initiate a pod restart through a causal chain.

\subsection{Import/Export}

SPL provides a pub/sub mechanism between jobs in the same instance through the
\code{Import} and \code{Export} operators~\cite{spl_import, spl_export}. These
operators allow users to construct microservices out of their applications: they
are loosely connected, can be updated independently and the platform is
responsible for resolving subscriptions. A common pattern we have seen in
production is users will deploy an ingest application for first-level parsing.
It publishes tuples through an \code{Export} operator, and various analytic
applications subscribe via their \code{Import} operators. The ingest application
always runs, while the analytic applications can vary from always running
to quick experiments.
 
The three actors in this pub/sub system are:
\begin{enumerate}
    \item \code{Export} operator. Publishes its input stream through a name 
        or a set of properties.
    \item \code{Import} operator. Subscribes to a stream based on its name or 
        a set of properties. Stream content can also be filtered on tuple 
        attributes using a filter expression.
    \item Subscription broker. Part of the platform, it's responsible for discovering 
        matches between \code{Import} and \code{Export} operators during job 
        submission and notifying PEs to establish new connections.
\end{enumerate}

\paragraph{Legacy:} Upon job submission, the platform creates states for all
\code{Import} and \code{Export} operators found in the job's graph metadata and
stores them in ZooKeeper. It then invokes the subscription broker to compute new
available routes and send route update notifications to the relevant PEs.  Users
can modify subscriptions at runtime either programmatically in an application
through SPL and native language APIs, through a command line tool, or through
dashboards.  Changes are relayed to the subscription broker through the platform
using the JMX protocol. Upon reception of such modifications, the subscription
broker reevaluates possible import and export matches and sends
route updates to the relevant PEs.

\paragraph{Cloud native:} We represent \code{Import} and \code{Export} operators
as CRDs.  During job submission, each instance of such an operator in an
application becomes a separate CRD. Users can update subscription properties by
editing the CRD itself. The subscription broker is a conductor that
observes events on both import and export CRDs. It maintains a local
subscription board, and when it discovers a match, it notifies the relevant PE.
Note that this subscription board is local state that can be lost: upon restart,
the subscription broker will reconstruct it based on re-receiving all
notifications from Kubernetes.  The PEs ignore any redundant subscription
notifications.

We replaced the JMX interface with a REST service endpoint (see
\S~\ref{sec:coupling}), periodically polled by the PEs to watch for
changes. We replace the synchronous JMX notification with a loosely coupled UDP
notification from the subscription broker to the PEs.  Alterations to the
import and export states from the application are also done
through the REST service. This service and the import and export
controllers are concurrent agents. To avoid race conditions, the REST service
uses the import and export coordinators for state changes.

\subsection{Consistent Regions}
\label{sec:consistentregion}

Streams provides application-level fault-tolerance through \emph{consistent
regions}~\cite{consistent_regions}. A consistent region is a region of an
application which guarantees at-least-once tuple processing.  The \emph{job
control plane} (JCP) periodically coordinates a consistent checkpointing protocol
where operators checkpoint their local state upon seeing special punctuations in their
streams. The checkpoints are stored in highly available external storage, such
as RocksDB or Redis. The JCP is composed of:

\begin{enumerate}
    \item A job-wide coordination system that orchestrates the consistency
        protocol across the job's consistent regions using a finite-state machine.
    \item A runtime interface embedded in each PE that interacts with the 
        coordination system in the JCP.
\end{enumerate}

When a PE fails or a PE-to-PE connection drops, the JCP initiates
rollback-and-recovery: failed PEs restart, all PEs instruct their operators to
rollback state to the last known-good checkpoint, and sources resend all tuples
whose resultant state was lost during the rollback.

\paragraph{Legacy:} The JCP coordination system is implemented as an SPL
operator~\cite{JCPOperator} with a Java backend and uses a JMX message bus.
Once instantiated, this operator registers itself with the platform as a JMX
service endpoint. Similarly, PEs that are part of a consistent region bootstrap
their JCP runtime interface at startup.  This interface also registers itself as
a JMX endpoint with the platform.  The platform is a message broker between the
JCP coordination system and the runtime interfaces of the PEs.  Checkpointing is
configured at the job level and the configurations are pushed to the PEs during
instantiation. Each checkpointing option has a bespoke configuration system that
must be determined manually by the Streams user or administrator. Life cycle
events, such as PE failure, are handled by the platform and forwarded to the JCP
coordination system.  Lastly, the coordination system implements its own fault
tolerance by storing its internal state in ZooKeeper. The storage configuration
must be determined by the user and manually set as an application parameter. For
instance, when using Redis, users must manually specify the names of shards and
replication servers~\cite{streamsredis}.

\paragraph{Cloud native:} We did not change the consistent region protocols, as
they are application-level. We also did not try to use Kubernetes CRDs to store
operator checkpoints: they will be of an arbitrary number and size, and we
wanted to maintain a clear separation between platform and application concerns.

However, we did address architectural inefficiencies by applying the loose
coupling principles (\S~\ref{sec:coupling}).  We moved the JCP coordination
system into its own Kubernetes operator, which avoids making the instance
operator the message broker between the JCP runtime and coordination system. At
submission time, the instance operator creates a consistent region operator for
each consistent region in a job. The consistent region operator monitors
resource events through controllers and conductors. It also manages its own CRD,
\code{ConsistentRegion}, used to persist internal states.

We no longer rely on a JMX message bus because Kubernetes serves that purpose:
controllers and conductors receive resource event notifications from Kubernetes.
The consistent region operator subscribes to pod life cycle events, PE
connectivity and consistent region state change events.  In the current version
we use a REST service to propagate consistent region changes to PEs. PEs also
use this service to notify both the instance operator and the consistent region
operator of connectivity and consistent region state changes.

As an example of our composability design goal (\S~\ref{sec:arch}), we
automatically configure checkpoint storage. To use Redis with cloud native
Streams, users specify a Redis cluster's service name.  The instance operator
discovers all available servers in that cluster through the service's DNS record
and automatically computes the appropriate configuration.

\subsection{System Tests}

Streams relies on over 2,400 application system tests for the development and
release cycle. Accumulated from a decade of product development, each test uses
at least one SPL application. The tests cover integration and regression testing
for all core Streams application features. They are split into two major
categories: those which require the distributed platform, and those that do not.
The tests which do not require the distributed platform execute the application
in a single PE, running in a normal Linux process. These tests primarily focus
on the correctness of the compiler, application runtime semantics, and operators
from the standard library. The tests which do require the distributed platform
test many of the application features covered in this paper: PE-to-PE
communication, metrics, consistent regions, parallel regions and any sort of
application behavior which requires non-trivial interaction with its external
environment. The test which require the distributed platform use multiple PEs (up 
to hundreds) and some use multiple SPL applications.

Tests are organized in scenarios containing a list of steps to perform,
environment variables to use, and context tags to honor. Context tags are
descriptors used by the test harness driver to determine the appropriate node
the test must run on, attributes such as the operating system version or whether
the node must be equipped with a network accelerator.

A variety of test steps are available, ranging from moving files around to
randomly killing critical processes. Test success or failures are determined
using special steps called probes that wait for the system to reach a particular
state to complete, such as waiting for a job to be in the \code{Submitted}
state, or waiting for all the processing elements of a job to be in the
\code{Connected} state.

\paragraph{Legacy:} In order to operate with legacy Streams, our system test
framework must be pre-installed on the target cluster. The target node names
must be known and collected in the framework's configuration files. The cluster
must also have a specific file system layout and sharing configuration in order
for tests expecting shared files to operate. The version of IBM Streams
being tested must be available at the same place in the file hierarchy to all
nodes in the cluster. If the test application writes or reads files, those 
files must be available over a Network File System.

\paragraph{Cloud native:} Similar to our platform itself, we organize the system
we use to test it around Kubernetes operators and CRDs. We define a
\code{TestSuite} CRD which maintains five lists of tests: pending, running,
passed, failed and aborted. It also maintains testing parameters such how to
select tests to run, how many tests to run concurrently, how many failures to
tolerate before stopping a run, and what to do with testing artifacts.  Users
initiate a new test run by creating a \code{TestSuite} CRD which specifies which
tests to run. The \code{TestSuite} controller will select $N$ tests to go on the
running list, where $N$ is the concurrency number. The remaining tests that meet
the selection criteria go on the pending list. The \code{TestSuite} controller
then creates a pod for each test on the running list.  When a test pod finishes,
the pod controller uses a \code{TestSuite} coordinator to indicate test success
or failure. The coordinator computes how the test lists should be modified,
creates a new pod for the next test on the pending list, and finally updates the
CRD itself to match the computed test lists.  This process repeats until the
pending list is empty, or the failed and aborted list exceed the failure
threshold.

The \code{TestSuite} controllers run in a test harness Kubernetes operator. This
test harness architecture enjoys all the benefits of being cloud native. It can
run on any Kubernetes cluster; it does not require any system-specific
configuration except node labels to expose available hardware accelerators; it
makes testing for a specific operating system version irrelevant as both cloud
native Streams and the test framework are distributed as container images; it
makes test run completely discoverable through the use of the \code{TestSuite}
CRD; it is resilient to failures because all important state is stored in the
CRD; and test runners and test executions can be monitored like any other pods
in the cluster with standard tools like Prometheus and Grafana. Finally, the
harness operator is blind to the kind of test runners it creates as from its
perspective it only manipulate pods and their execution states.

\section{Lessons learned}
\label{sec:lessons}

During design, implementation and testing, we adopted lessons that
served as general guidance:

\begin{enumerate}
    \item \emph{Don't store what you can compute.} Storing state in a 
        distributed system is expensive---not just in bytes and bandwidth, 
        but in complexity. Modifying that state requires transactions and forces
        components to synchronize. This complexity will necessarily infect the 
        rest of the system. If it is possible to recompute a result, the cost in 
        cycles buys a simpler design.
    \item \emph{Align your design with Kubernetes concepts.} Alignment enables
        integration and simplification. We did not have to implement any
        management of Streams instances in cloud-native Streams because we
        enforce one Streams instance per Kubernetes namespace. In legacy
        Streams, a domain managed multiple instances. We get that for free as
        our ``domain'' just becomes the Kubernetes cluster.
    \item \emph{Don't re-implement what Kubernetes already provides.} The value 
        in using a general purpose distributed platform is not having to 
        re-implement the basics. Designs which require implementing bespoke 
        versions of life cycle management, communication, storage or 
        configuration not only waste code and effort. Such designs are also less 
        likely to integrate well into Kubernetes, forcing even more bespoke 
        implementations of other features.
    \item \emph{Rely on Kubernetes for atomicity, consistency and redundancy.} 
        Kubernetes provides reliable storage, and sends totally ordered, reliable 
        notifications based on changes to the objects in that storage. Building 
        systems using these primitives allows for simpler, better integrated 
        designs.
    \item \emph{Always use hierarchical, deterministic naming.} Globally unique 
        names in a distributed system are a form of state: creating them requires 
        synchronization to avoid duplicates, and their metadata must be durably 
        stored. For top-level objects in a system, this property is 
        unavoidable. But named objects nested in those top-level objects do not
        need to be globally unique, as their top-level object is an implicit 
        namespace. Requiring such nested names to be globally unique imposes 
        unnecessary state management and synchronization. Hierarchical, 
        deterministic naming schemes allows other entities in the system to 
        \emph{compute} what the names must be.
\end{enumerate}

\section{Results}
\label{sec:results}

\subsection{Experimental results}

Raw performance was not a motivation (\S~\ref{sec:motivation}) or design goal
(\S~\ref{sec:arch}). However, if cloud native Streams' platform performance was
significantly worse than legacy Streams', then it would not be an acceptable
replacement. The primary goal of our experiments is to demonstrate that cloud
native Streams has acceptable comparable performance, and the secondary goal is
to identify aspects of Kubernetes which can be improved.

We ran our experiments on a 14 node cluster using Streams v4.3.1.0 as legacy,
Kubernetes v1.14 and Docker v18.09.6. Each node has two 4-core Intel Xeon X5570
processors at 2.93 GHz with hyperthreading enabled and 48GB of memory.  One
node is dedicated to management, leaving 13 nodes and 104 physical cores (208
logical cores) for applications.

Unless otherwise stated, our test application has a source operator which
continuously generates tuples and feeds into an $n$-way parallel region. Each
channel in the parallel region has a pipeline of $n$ operators, and all
channels eventually converge into a sink operator. We fuse each operator into
its own PE. We vary $n$ in our experiments, which means that the number of
operators and PEs grows with $n^2$. As described earlier, each PE is a separate
process and runs in its own pod. Different experiments need a different number
of pre- and post-processing operators before and after the parallel region.

\begin{figure}
\centering
\begin{subfigure}{1.0\linewidth}
  \centering
  \includegraphics[width=\linewidth]{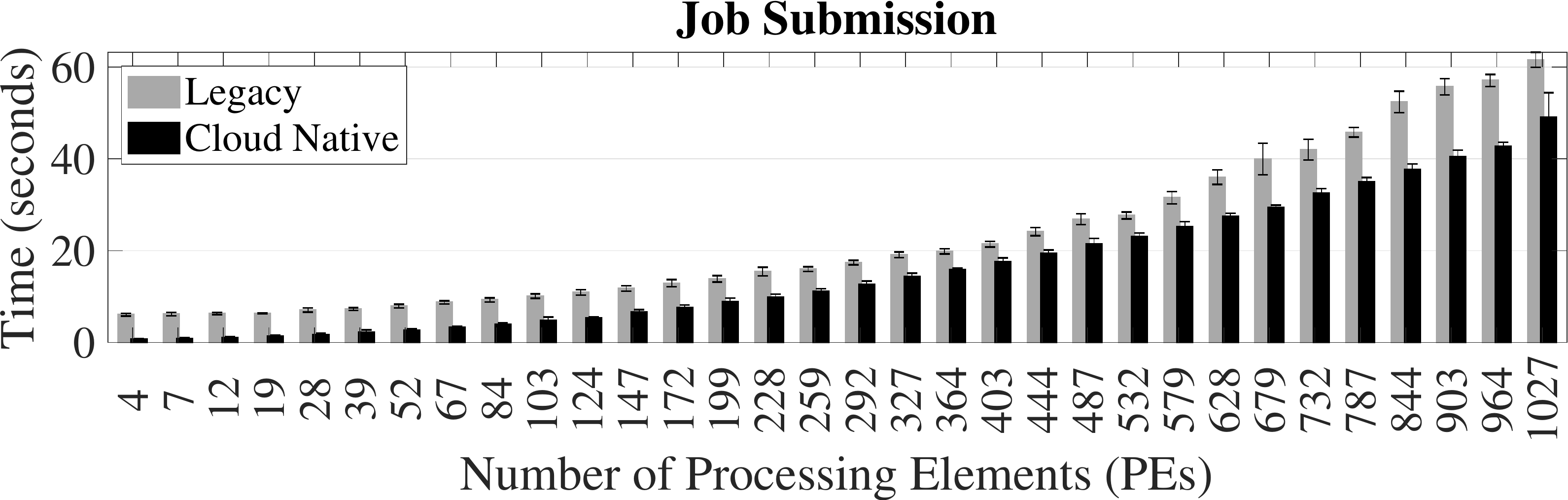}
  \caption{Time taken for jobs to reach the \code{Submitted} state.}
  \label{fig:submit-cancel-submitted}
\end{subfigure}
\begin{subfigure}{1.0\linewidth}
  \centering
  \includegraphics[width=\linewidth]{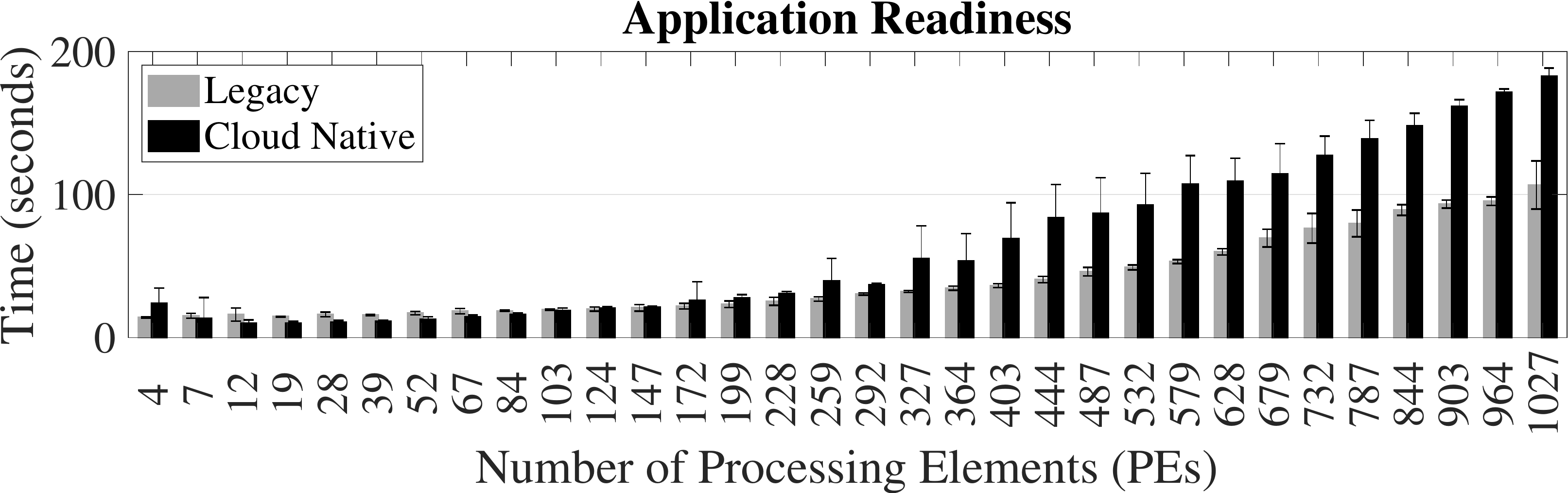}
  \caption{Time taken for jobs to reach full health.}
  \label{fig:submit-cancel-healthy}
\end{subfigure}
\begin{subfigure}{1.0\linewidth}
  \centering
  \includegraphics[width=\linewidth]{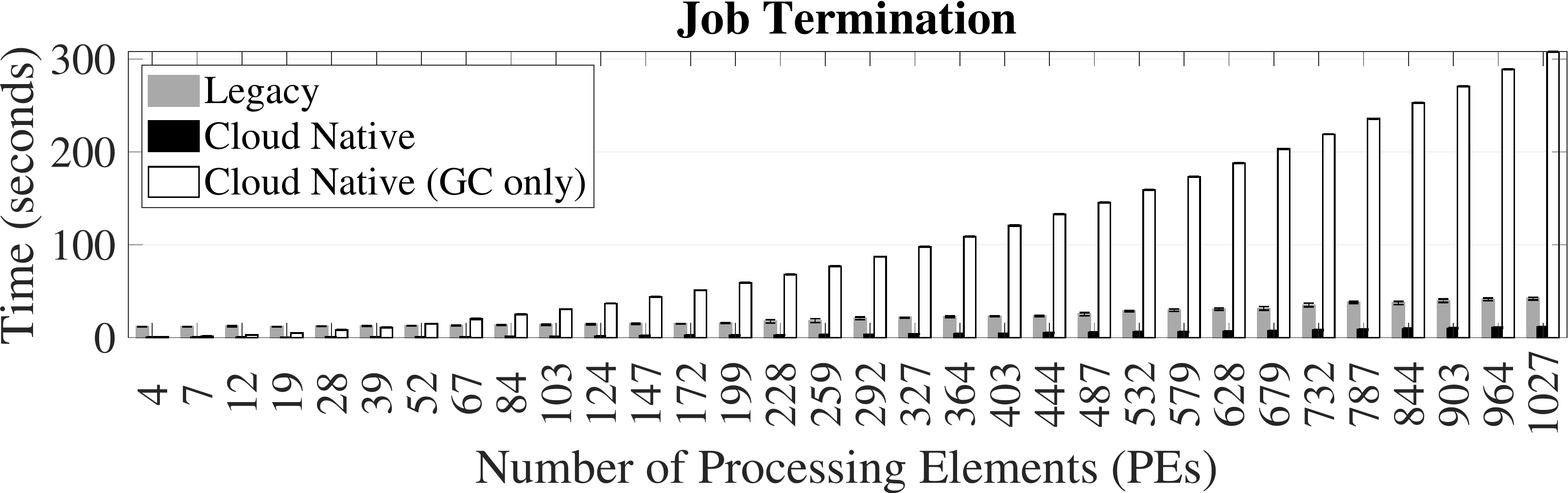}
  \caption{Time taken for jobs to fully terminate.}
  \label{fig:submit-cancel-terminate}
\end{subfigure}
\caption{Job life cycle times.}
\label{fig:submit-cancel}
\end{figure}

\paragraph{Job life cycle:}
\label{sec:results:job-submission-termination}

The three job life cycle phases exposed to users are submission, full health
and full termination. The submission time is how long it takes for the
platform to create all of the PEs and job resources. Such jobs are still
initializing and are not yet processing data. Only when all PEs are running and
have established all connections is the application processing data and
considered fully healthy. Finally, after the user cancels the job, the platform
considers the job fully terminated after all PEs and associated resources are
gone.

Figure~\ref{fig:submit-cancel} shows how long it takes to reach each phase of
the job life cycle for both cloud native and legacy Streams, with each data
point representing the average of 10 runs.
Figure~\ref{fig:submit-cancel-submitted} shows that cloud native Streams is
consistently faster to reach the \code{Submitted} state. Reaching the
\code{Submitted} state only requires resource creation; for legacy Streams, that
means registering all resources in ZooKeeper, and for cloud native Streams that
means all resources are stored in etcd. However, legacy Streams also
computes PE schedules: it rejects jobs for which it cannot find a valid
schedule. In cloud native Streams, Kubernetes schedules PE’s pods asynchronously.

The time it takes for cloud native Streams to reach full health,
Figure~\ref{fig:submit-cancel-healthy}, is dependent on whether the
cluster is oversubscribed. Each PE is a process in a separate pod, and the
experiments scale to 1027 PEs. But the cluster will be fully subscribed by at least
208 PEs; there are more processes than cores. Before the cluster is
fully subscribed, cloud native Streams performs competitively with legacy
Streams. Both versions suffer as the cluster becomes more oversubscribed, but
cloud native Streams eventually takes twice as long.

The job termination experiments, Figure~\ref{fig:submit-cancel-terminate}, have
results for two different approaches for cloud native Streams: manual resource
deletion and relying on Kubernetes' garbage collector. In the case of manual
deletion, the job controller actively cleans up by telling Kubernetes to delete
resources in bulk by their label. Bulk deletion minimizes the number of API calls
and therefore reduces the strain on Kubernetes' API server. We originally relied
on Kubernetes' resource garbage collector to automatically reclaim resources
owned by deleted jobs. Kubernetes' garbage collector, however, does not scale
well as the number of resources grows.

\begin{figure}[t]
    \centering
    \includegraphics[width=1.0\linewidth]{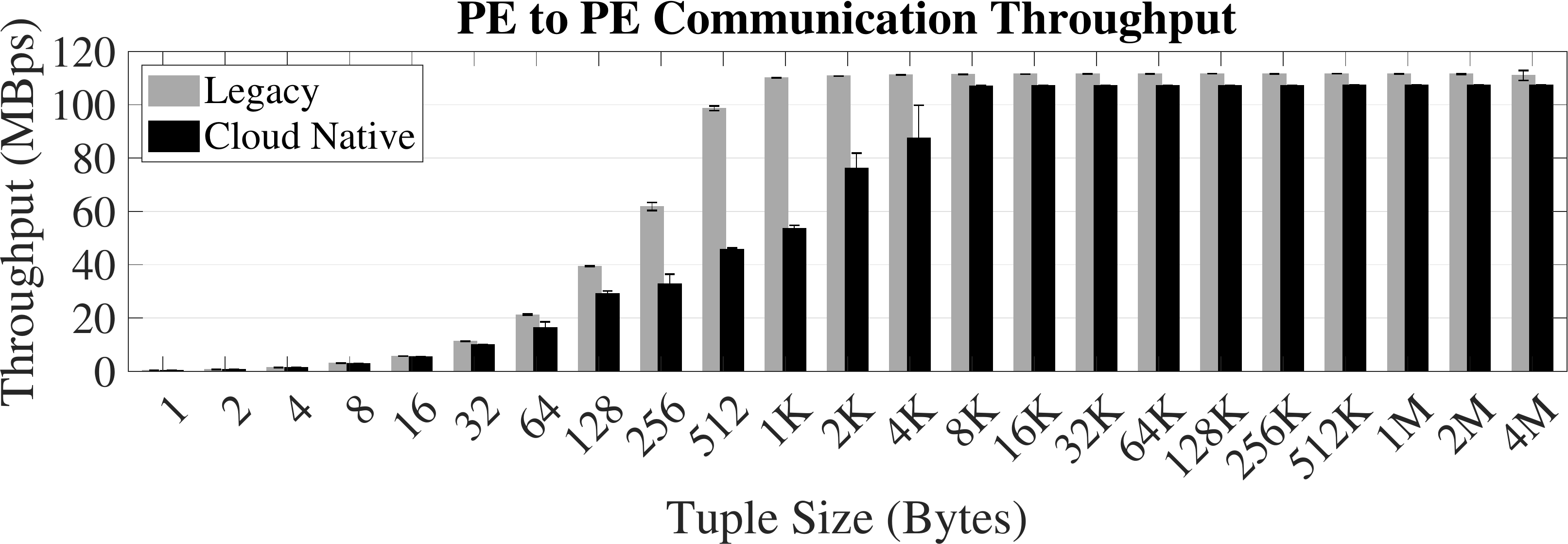}
    \caption{PE-PE communication throughput.}
    \label{fig:throughput}
\end{figure}

\paragraph{PE-to-PE communication throughput:} The experiments in
Figure~\ref{fig:throughput} use an application designed to test PE-to-PE
throughput. It consists of two PEs, pinned to two fixed nodes in our
testbed, while varying the size of the tuple payload from 1 byte to 4 MB.
Transmissions run for 5 minutes, with throughput measurements every 10 seconds.

Cloud native Streams achieves significantly lower throughput than legacy when
the tuple size is smaller than 4 KB. This performance degradation is due to the
deeper networking stack used within Kubernetes. Because of its networking
architecture, a single packet sent from one container is required to cross
various virtual interfaces and packet filters before reaching another container.
Comparatively, a packet sent between two PEs within the legacy Streams platform
is directly sent to the default interface for the target route. This increased
complexity has the most pronounced effect with payloads less than 8 KB.  With
larger payloads, the increased networking cost is mostly amortized.

\begin{figure}[t]
    \centering
    \includegraphics[width=1.0\linewidth]{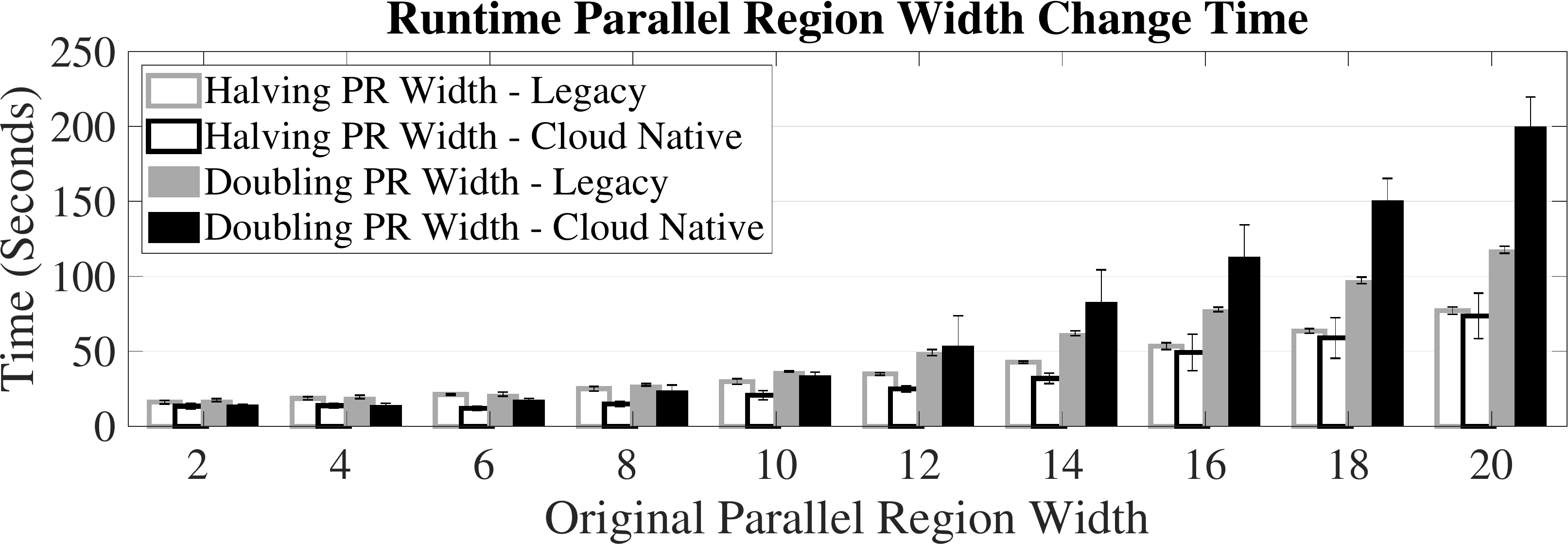}
    \caption{Parallel region width change time.}
    \label{fig:pr-width-change}
\end{figure}

\begin{figure}[t]
    \centering
    \includegraphics[width=\linewidth]{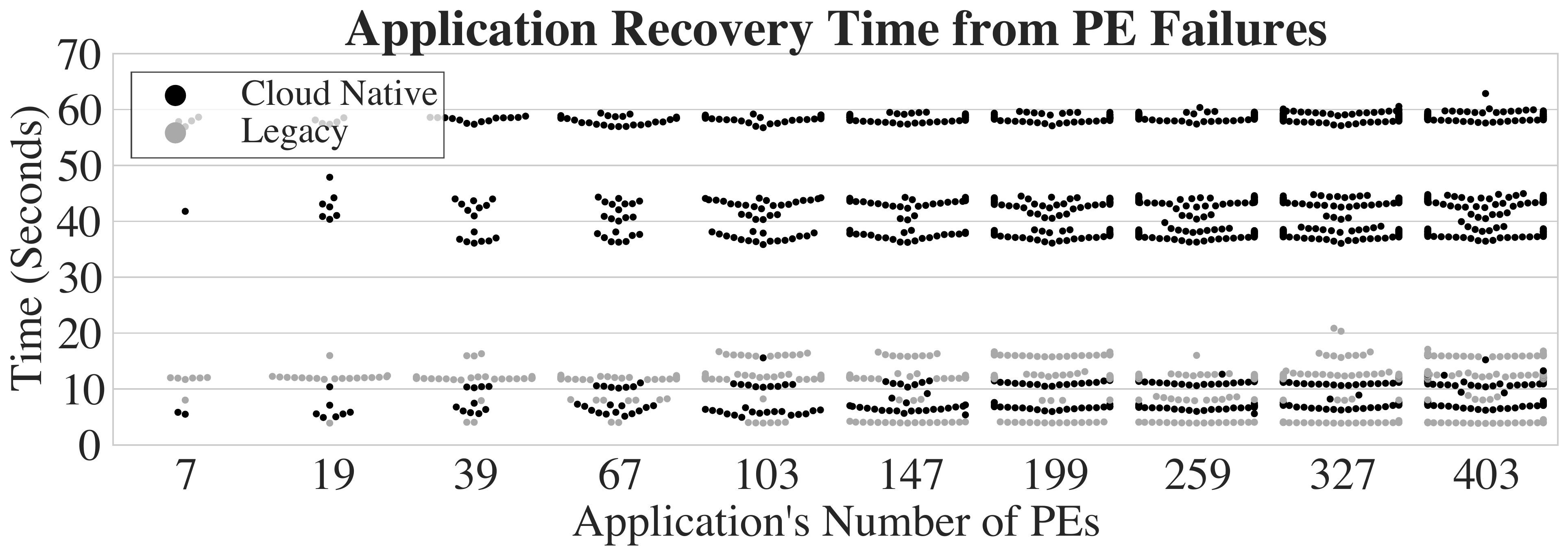}
    \caption{Application PE failure recovery time.}
    \label{fig:pe-recovery}
\end{figure}

\begin{figure}[t]
    \centering
    \includegraphics[width=1.0\linewidth]{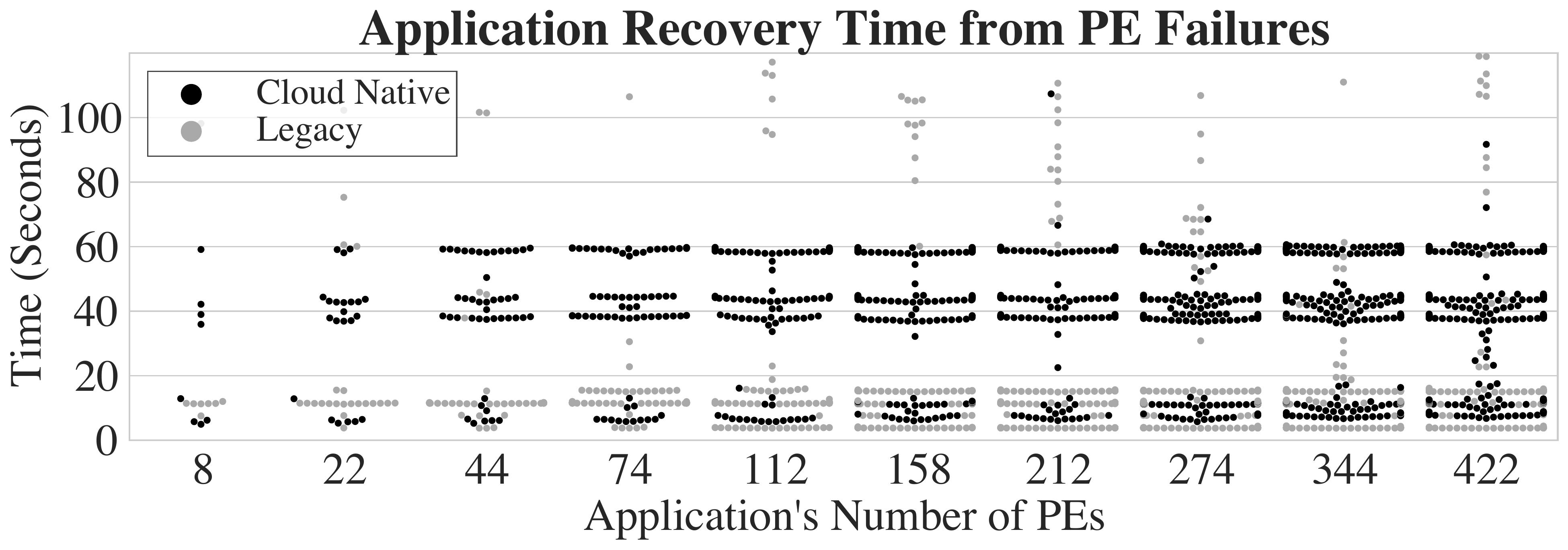}
    \caption{CR App. PE Failure Recovery.}
    \label{fig:cr-pe-recovery}
\end{figure}

\paragraph{Parallel region width change:}

Figure~\ref{fig:pr-width-change} has two sets of experiments: increasing and
decreasing parallel region width. The $x$-axis is the starting width of the
parallel region, and we measure how long it takes an application currently at
full health to either double or halve that parallel region. In general, cloud
native Streams benefits from the concurrency enabled by the design from
\S~\ref{sec:parupdates}. Starting new PEs and terminating the old PEs
happens concurrently, handled by Kubernetes' pod scheduling. In legacy Streams,
these phases need to happen sequentially. Cloud native Streams also has less
communication with its metadata backing store (etcd) than legacy Streams
(ZooKeeper). Starting at a width of 12, the cluster is oversubscribed (the
number of PEs grows with the square of the width). At that point, when doubling
the width, we see the same behavior regarding starting new PEs in an oversubscribed
cluster as in Figure~\ref{fig:submit-cancel-healthy}.

\paragraph{PE failure recovery:}
\label{sec:results:pe-recovery}

We test PE recovery time in Figure~\ref{fig:pe-recovery}. The $x$-axis is the
number of PEs in the application, and each dot represents how long it took the
application to return to full health after we killed a particular PE. This
process times how long it takes for the platform to detect that the PE is gone,
start a replacement, and wait for the replacement PE to re-establish all
connections. The clustered times are due to different PEs being in similar
places in the application topology. In cloud native Streams, the delay is in
re-establishing the connections; the PEs are restarted quickly as that is almost
entirely handled through Kubernetes' pod management.

\paragraph{Consistent region PE failure recovery:}

The experiments in Figure~\ref{fig:cr-pe-recovery} are the same as in the PE
failure recovery with the addition that the operators are also in a consistent
region. This addition means that their recovery is managed by the consistent
region protocol (\S~\ref{sec:consistentregion}) which requires more
communication and coordination than just restarting the PEs.  The outlier
latencies tend to be the PEs which have more connections.

\paragraph{Discussion:}

One benefit of a bespoke platform is specialization. Our experimental results
show that there is currently a cost for some actions when using Kubernetes as
a generic platform. Improving these parts of Kubernetes will improve
its ability to handle workloads such as Streams.

\emph{Oversubscription:} Cloud native Streams behaves poorly compared to legacy
Streams when the cluster is oversubscribed.  We have identified two potential
culprits.  First, the DNS propagation in Kubernetes seems to be slower than the
name resolution mechanism in legacy Streams. This latency is likely caused by
the extra complexity of pod networking. Second, many more subsystems are
involved in cloud native Streams than legacy Streams when creating new PEs:
where \code{fork()} is enough for the latter, the former calls upon the Docker
daemon and various Linux kernel facilities such as \code{cgroups} to start new
pods.

\emph{Networking latency:} The increased latency is the biggest pain point as
IBM Streams was designed and engineered as a low latency, high throughput
streaming solution. This is especially true as the most common tuple size used
by Streams customers is around 500 bytes, within the size range where the
latency degradation reaches 50\%. A solution is to use two different planes for
control and data: Kubernetes networking for the Streams control plane, while a
separate network for the Streams data plane. Such a separation can be achieved
through user-space networking, either through a bespoke user-space TCP/IP
network stack integrated into Streams' runtime or through a Kubernetes plugin
supporting user-space networking, such as Microsoft FreeFlow\cite{freeflow}.

\emph{Garbage collector:} When handled completely by the Kubernetes garbage
collector, our resource deletion time experienced significant latency even with
a modest number of resources on an undersubscribed cluster. The garbage
collector could likely be tuned to reduce the deletion time, but such tuning
introduces the danger of overfitting it to one specific workload at the
detriment of others. Garbage collector plugins similar to scheduler plugins
could solve this problem.

\emph{PE recovery:} We initially suspected the container runtime added latency,
but further investigation conducted by stressing container creation and
deletion did not show any behavior that would explain the increasing recovery
latency past 100 PEs. Another intuition concerns the networking address
allocation for PEs: when recovering a failed PE, the legacy Streams platform
will respawn the process on the same host. By doing so, the name resolution
stays stable and other PEs communicating with the respawned process will be
able to reconnect to it immediately. However, on the Kubernetes platform, PEs
may not end up with the same container IP address, even when allocated on the
same host. Therefore, all PEs communicating with the respawned process first
need to get a fresh name translation record, which is dependent on how fast the
Kubernetes DNS subsystem propagate changes.  Validation requires more
investigation. However, some workloads may benefit from stable IP addresses
for pods. Such stable addressing could be implemented by either updating an
existing network plugin or implementing a network plugin specific to the
workload.

\subsection{Lines of Code}

Rearchitecting a legacy product to be cloud native should offload significant
responsibility to the cloud platform. This process should significantly
reduce the lines of code in the implementation. Table~\ref{tab:sloc} shows
that reduction.

\begin{table}[h]
    \centering
    \footnotesize
    \begin{tabular}{|l|r|r|}
        \hline
                             & \textbf{legacy}   & \textbf{cloud native}  \\
        \hline
        \textbf{SPL}         &           429,406 &                415,809 \\
        \textbf{platform}    &           569,933 &                148,375 \\
        \textbf{install}     &            14,785 &                      0 \\
        \hline
        \hline
        \textbf{total}       &         1,014,124 &                564,184 \\
        \hline
    \end{tabular}
    \caption{Physical lines of source code across Streams versions.}
    \label{tab:sloc}
\end{table}

We use scc~\cite{scc} to count code. It counts physical source lines of code,
which is defined as lines of code which are not comments or blank. The
languages included in our count are C++, Java, Perl, XML Schema and YAML.  The
SPL compiler is primarily C++, with some of the user-exposed code generation
features in Perl.  The SPL runtime is split between C++ and Java. The legacy
platform is about 80\% Java, 20\% C++. The install is primarily Java.  We do
not include code related to the build process or system tests.

Cloud native Streams is about half the size of the legacy version, and the
platform is about a quarter the size of the old platform. The implementation of
the architecture presented in this paper is about 26,000 lines of code, which
means that we reused about 122,000 lines of platform code. Most of that code is
the job submission pipeline (\S~\ref{sec:jobsubmission}). The cloud native
version does not have an installer: users apply the YAMLs for the CRDs and make
the Docker images available which contain the instance operator and the
application runtime.

\section{Conclusion}

Cloud native Streams replaces the IBM Streams platform with Kubernetes. It
offloads life cycle management, scheduling, networking and fault tolerance to
Kubernetes. It does this by using Kubernetes as a state-management-service: all
important state is managed by Kubernetes, and its services react to state change
events delivered by Kubernetes. Those services implement the cloud native
patterns presented here: controllers, conductors, coordinators and the causal
chains formed by their interactions.  Other workloads can use these patterns to
implement their own cloud native platforms. We have also experimentally
demonstrated areas in which Kubernetes needs improvement to better serve as a
generic platform: performance in an oversubscribed cluster, networking latency,
garbage collector performance and pod recovery latency.

\clearpage
\pagebreak

\bibliographystyle{unsrt}
\bibliography{streams_knative}

\end{document}